\documentclass[prl,nofootinbib,floatfix,twocolumn,showpacs,superscriptaddress]{revtex4}
\usepackage{graphicx,amsmath}% Include figure files
\usepackage{dcolumn}% Align table columns on decimal point
\usepackage{bm}% bold math
\usepackage{rotating}
\usepackage{amsmath,amssymb,amsfonts}
\usepackage{color}

\def\pythia6{{\sc Pythia6~}}
\def\pythia{{\sc Pythia~}}

\begin{document}

\title{Transverse momentum broadening of hadrons produced in semi-inclusive deep-inelastic scattering on nuclei.}  
%  \underline{Paper Tag:}
%  \hspace*{1cm}\underline{Drafting Committee: 70}}

% List of Institute Addresses 

\def\groupargonne{\affiliation{Physics Division, Argonne National Laboratory, Argonne, Illinois 60439-4843, USA}}
\def\groupbari{\affiliation{Istituto Nazionale di Fisica Nucleare, Sezione di Bari, 70124 Bari, Italy}}
\def\groupbeijing{\affiliation{School of Physics, Peking University, Beijing 100871, China}}
\def\groupcolorado{\affiliation{Nuclear Physics Laboratory, University of Colorado, Boulder, Colorado 80309-0390, USA}}
\def\groupdesy{\affiliation{DESY, 22603 Hamburg, Germany}}
\def\groupzeuthen{\affiliation{DESY, 15738 Zeuthen, Germany}}
\def\groupdubna{\affiliation{Joint Institute for Nuclear Research, 141980 Dubna, Russia}}
\def\grouperlangen{\affiliation{Physikalisches Institut, Universit\"at Erlangen-N\"urnberg, 91058 Erlangen, Germany}}
\def\groupferrara{\affiliation{Istituto Nazionale di Fisica Nucleare, Sezione di Ferrara and Dipartimento di Fisica, Universit\`a di Ferrara, 44100 Ferrara, Italy}}
\def\groupfrascati{\affiliation{Istituto Nazionale di Fisica Nucleare, Laboratori Nazionali di Frascati, 00044 Frascati, Italy}}
\def\groupgent{\affiliation{Department of Subatomic and Radiation Physics, University of Gent, 9000 Gent, Belgium}}
\def\groupgiessen{\affiliation{Physikalisches Institut, Universit\"at Gie{\ss}en, 35392 Gie{\ss}en, Germany}}
\def\groupglasgow{\affiliation{Department of Physics and Astronomy, University of Glasgow, Glasgow G12 8QQ, United Kingdom}}
\def\groupillinois{\affiliation{Department of Physics, University of Illinois, Urbana, Illinois 61801-3080, USA}}
\def\groupmichigan{\affiliation{Randall Laboratory of Physics, University of Michigan, Ann Arbor, Michigan 48109-1040, USA }}
\def\groupmoscow{\affiliation{Lebedev Physical Institute, 117924 Moscow, Russia}}
\def\groupnikhef{\affiliation{National Institute for Subatomic Physics (Nikhef), 1009 DB Amsterdam, The Netherlands}}
\def\groupstpetersburg{\affiliation{Petersburg Nuclear Physics Institute, Gatchina, Leningrad region, 188300 Russia}}
\def\groupprotvino{\affiliation{Institute for High Energy Physics, Protvino, Moscow region, 142281 Russia}}
\def\groupregensburg{\affiliation{Institut f\"ur Theoretische Physik, Universit\"at Regensburg, 93040 Regensburg, Germany}}
\def\grouprome{\affiliation{Istituto Nazionale di Fisica Nucleare, Sezione Roma 1, Gruppo Sanit\`a and Physics Laboratory, Istituto Superiore di Sanit\`a, 00161 Roma, Italy}}
\def\grouptriumf{\affiliation{TRIUMF, Vancouver, British Columbia V6T 2A3, Canada}}
\def\grouptokyo{\affiliation{Department of Physics, Tokyo Institute of Technology, Tokyo 152, Japan}}
\def\groupamsterdam{\affiliation{Department of Physics and Astronomy, Vrije Universiteit, 1081 HV Amsterdam, The Netherlands}}
\def\groupwarsaw{\affiliation{Andrzej Soltan Institute for Nuclear Studies, 00-689 Warsaw, Poland}}
\def\groupyerevan{\affiliation{Yerevan Physics Institute, 375036 Yerevan, Armenia}}
\def\groupnone{\noaffiliation}

% Set Institute Order 

\groupargonne
\groupbari
\groupbeijing
\groupcolorado
\groupdesy
\groupzeuthen
\groupdubna
\grouperlangen
\groupferrara
\groupfrascati
\groupgent
\groupgiessen
\groupglasgow
\groupillinois
\groupmichigan
\groupmoscow
\groupnikhef
\groupstpetersburg
\groupprotvino
\groupregensburg
\grouprome
\grouptriumf
\grouptokyo
\groupamsterdam
\groupwarsaw
\groupyerevan

% List of Authors 

\author{A.~Airapetian} \groupgiessen \groupmichigan
\author{N.~Akopov}  \groupyerevan
\author{Z.~Akopov}  \groupdesy
\author{E.C.~Aschenauer}  \groupzeuthen
\author{W.~Augustyniak}  \groupwarsaw
\author{A.~Avetissian}  \groupyerevan
\author{E.~Avetisyan}  \groupdesy
\author{B.~Ball}  \groupmichigan
\author{S.~Belostotski}  \groupstpetersburg
\author{N.~Bianchi}  \groupfrascati
% ...

\collaboration{The {\sc Hermes} Collaboration} \noaffiliation

%\date{\today}

%{\center DRAFT version 5.62 - journal: PLB}

\begin{abstract}
%\newpage
%\internallinenumbers                                                   %LINE_NUMBERING
The first direct measurement of the dependence on target nuclear mass of the 
average squared transverse momentum $\langle p_t^2 \rangle$ of $\pi^+, \pi^-$, and $K^+$ mesons from deep-inelastic lepton scattering is obtained
as a function of several kinematic variables.
The data were accumulated at the {\sc Hermes} 
experiment at {\sc Desy}, in which the {\sc Hera} $27.6$~GeV
lepton beam was scattered off several nuclear gas targets.  
The average squared transverse momentum was clearly observed to increase
with atomic mass number. The effect increases as a function of $Q^2$ and $x$ and remains constant 
as a function of both the virtual photon energy $\nu$ and the fractional hadron energy $z$, except that it vanishes
as $z$ approaches unity.
\end{abstract}


\maketitle
%\runninglinenumbers    %LINE_NUMBERING
The evolution of a fast-moving quark into
hadrons is a non-perturbative and dynamic phenomenon.
The basic process of hadronization in vacuum is 
described by a well-developed phenomenology, 
constrained primarily by data from inclusive hadron production in $e^+e^-$
annihilation. More recently, semi-inclusive
hadron multiplicities 
measured in deep-inelastic
scattering (DIS) of leptons
on protons and deuterons, 
together with inclusive hadron yields from the {\sc Rhic} p-p collider experiments, 
have provided further constraints. For a recent global analysis of such data, see Ref.~\cite{FSS} and references therein.

The 
nuclear modification of 
hadron production in DIS 
was first observed by the pioneering semi-inclusive DIS experiments at {\sc Slac} \cite{Slac},
followed by measurements performed by the {\sc Emc} \cite{EMC} and the E665 \cite{E665} 
collaboration. 
These experiments typically determined the ratio of hadron multiplicities
observed in the scattering on a nucleus to those on Deuterium (D),
the so-called nuclear attenuation.
More recently, 
much more precise data were collected and analyzed by the 
{\sc Hermes} collaboration \cite{herm1}-\cite{herm4} as a function of the kinematic variables $\nu$, $z$, $Q^2$, and $p_t^2$, 
where $\nu$ is the energy of the virtual photon, $z$ the fractional hadron energy in the target rest frame, $-Q^2$ being the squared four-momentum of the virtual photon, and $p_t$ the transverse momentum of the produced hadron.
It has been suggested that hadron production proceeds through three qualitatively distinct stages
that involve the propagation and interaction of:

(i) the initial struck quark\footnote[1]{``quark'' is used in this Letter for both quarks and antiquarks} (the ``partonic'' stage), 

(ii) the subsequently formed
colorless state (the ``color-neutralization'' stage, often also termed 
``pre-hadronic'' stage), 
and  

(iii) the final produced hadrons (the ``hadronic'' stage).

This picture is supported by the results of a two-dimensional analysis of the multiplicity ratio \cite{herm4} in the kinematic variables 
$\nu $ and $z$. However, the existence and relative importance of the 
various stages
to the observed nuclear attenuation have been difficult to determine unambiguously.
This Letter reports, complimentary to the results presented in Ref. \cite{herm4}, the first detailed
measurement of another 
observable, which may help
to better constrain models, 
especially with regard to the role of the ``partonic'' stage: 
the broadening of the transverse momentum distribution of various hadrons.

In terms of the quark-parton model and QCD, there are several contributions to the
transverse momentum distribution of hadrons produced in semi-inclusive DIS:

(a) primordial transverse momentum,

(b) gluon radiation of the struck quark,

(c) the formation and soft multiple interactions of the ``pre-hadron'', and

(d) the interaction of the formed hadrons with the surrounding hadronic medium.

In semi-inclusive DIS off nuclear targets, the struck 
quark propagates through a ``cold'' nuclear medium. 
In the nuclear medium the primordial transverse momentum of quarks may be modified by various effects like Fermi motion of nucleons inside the nucleus, modification of the nucleon size, formation of non-nucleonic degrees of freedom like multiquark states 
or exchange mesons mediating the nuclear force. Also, the probabilities of the processes (b-d) may be enhanced resulting in a larger
transverse momentum magnitude of the observed hadrons relative to the process in the vacuum or in a free nucleon. 
In particular, process (b) may cause an increased transverse momentum magnitude and energy loss of the quark and therefore 
it has recently been suggested \cite{kopel2} that the 
broadening of hadron distributions in semi-inclusive DIS may be the most direct way to probe the ``partonic'' stage.
Both the radiative energy loss and the transverse momentum of the parton 
increase with $L$, which is the path length of the quark in the nuclear medium before the colorless state formation. QCD predictions for the relationship between energy loss and 
the broadening of $p_t$ distributions are given in Ref.
\cite{baier1,baier2}. As the relationship between these two 
quantities is independent of the dynamics of the initial scattering process, 
it holds equally well in ``cold'' nuclear matter and in finite-length ``hot'' matter, produced, 
{\it e.g.,} 
in ultra-relativistic heavy-ion collisions or high-energy proton-nucleus interactions. 
Thus 
the understanding of this broadening in the ``cold'' nuclear medium provides precious 
information for the interpretation of such high-energy processes
\cite{baier2,accardi}.

An indication of nuclear broadening of the transverse momentum distribution for charged hadrons
in muon and neutrino DIS was reported previously~\cite{EMC,herm4,SKAT} . 
This Letter presents the first direct measurement of 
the $p_t$-broadening for identified mesons and various nuclei as a function of the relevant kinematic variables.

The transverse hadron momentum $p_t$ is defined relative to the direction of the virtual photon, which is 
determined by the kinematics of 
the incident and the scattered lepton. 
The $p_t$-broadening $\Delta \langle p_t^2 \rangle_A^h $ is defined as the difference of the average squared transverse momentum 
of the detected hadron of type $h$ produced on a nuclear target with atomic mass number $A$ and that on a deuterium (D) target:
\begin{eqnarray}
\label{ptform}
\Delta \langle p_t^2 \rangle_A^h = \langle p_t^2 \rangle^h_A - \langle p_t^2 \rangle^h_D.   
\end{eqnarray}
The broadening may depend on the hadronic variable $z$, and on the leptonic variables $\nu$, $Q^2$, and $x$, with $x=\frac{Q^2}{2M\nu}$ the Bjorken variable and $M$ the nucleon mass. 

The measurements were performed at {\sc Hermes} using the $27.6$~GeV lepton ($e^+$ or $e^-$) beam stored in the {\sc Hera} ring at {\sc Desy}. The targets consisted of polarized or unpolarized D, unpolarized He, Ne, Kr, or Xe gas. The target atoms were injected into a 40~cm long, open-ended 
tubular storage cell through which the 
lepton beam passed. Target areal densities higher than 10$^{16}$~nucleons/cm$^{2}$ were obtained for unpolarized gases and of about $2 \cdot 10^{14}$ nucleons/cm$^{2}$ for polarized D. 
The scattered lepton and the produced hadrons were detected in coincidence by the {\sc Hermes} spectrometer \cite{herm}. Leptons were distinguished from hadrons with an average efficiency of 99\% and a contamination level of less than 1\% by using a transition-radiation detector, 
a scintillator pre-shower counter, a dual-radiator ring-imaging \v{C}erenkov ({\sc Rich}) detector, and an electromagnetic calorimeter. 
Scattered leptons were selected by imposing the constraints  
$Q^2 > 1~$GeV$^2$,  $W^2 > 10~$GeV$^2$, and $y < 0.85$,
where $W^2$ is the squared invariant mass of the photon-nucleon system and $y = \nu/E$ the energy fraction of the virtual 
photon in the target rest frame, with $E$ being the beam energy. 
In order to suppress the contributions from target remnant fragmentation, the requirement $z>0.2$ was imposed.

The numbers of identified and selected charged pions and positive kaons are given in Table~\ref{statistics}. Negative kaons 
and anti-protons 
are not considered in this analysis due to low statistical precision. Protons were not considered because they are not well enough described by the Monte Carlo simulation used for the unfolding procedure described below. 

\begin{table}
\begin{center}
\begin{tabular}{|c||c|c|c|c|}
\hline
Target &  $\pi^+ \times 10^3$ & $\pi^- \times 10^3$  & $K^+ \times 10^3$   \\
\hline
\hline
D   
& 1781 & 1445  
& 356   \\
\hline
He   
& 134 & 107   
& 27   \\
\hline
Ne   
& 380 & 303   
& 82   \\
\hline
Kr   
& 321 & 260   
& 72   \\
\hline
Xe  
& 193 & 157  
& 44   \\
\hline
\end{tabular}
\caption{ Accumulated yield of selected hadrons for the various nuclear targets.}
\label{statistics}
\end{center}
\end{table}
The data used for the analysis of $p_t$-broadening were corrected for 
QED radiative effects, instrumental resolution, and acceptance. This was achieved using a Monte Carlo simulation based unfolding procedure~\cite{unfolding}, which accounted for event migration between different kinematic bins. The {\sc Pythia} event 
generator \cite{pythia}, in conjunction with a special set of {\sc Jetset} fragmentation parameters \cite{hpythia}, was used together with the {\sc Radgen} \cite{radgen} simulation of QED radiative effects and with a {\sc Geant}~\cite{geant} based simulation of the detector. 
The possible dependence of the unfolding procedure on the target type used in the Monte Carlo simulation was studied with the {\sc Lepto} \cite{lepto} generator and {\sc Jetset} using D and Xe targets. The results for both targets were found to be consistent.
Therefore, a Monte Carlo simulation for the D target was used to correct the results for all 
targets. 

Selected events were binned in $p_t^2$ and in either $z$, $\nu$, $Q^2$, or $x$. These two-dimensional~\footnote{A two-dimensional unfolding still leaves room for acceptance effects if both the yield and the magnitude of the effect depends non-linearly on 
the variables that are not included in the two-dimensional binning. An unfolding in more than two dimensions was not feasible in this analysis.} distributions were unfolded using smearing 
matrices that hold information about migration among kinematic bins. This method inflates the statistical errors to account for this migration~\cite{andyunf}. The impact of the unfolding on $\langle p_t^2 \rangle$ ranges from 60\% to almost zero 
with increasing $\nu$, and is about 20\% for all $z$ bins. It varies from 20\% to 30\% with increasing $Q^2$, and is between 15\% and 40\% for the $x$-bins.
After this correction, the average $p_t^2$ value was calculated in each bin of $z$, $\nu$, $x$, or $Q^2$ and then $\langle p_t^2 \rangle_D$ was subtracted from $\langle p_t^2 \rangle_A$.  
The charged pion yields were corrected for pions from the decay of 
exclusively-produced $\rho^0$ mesons. A correction for such pions, where at least one decay pion was in the acceptance, was made using 
a special Monte Carlo generator \cite{rhomc} that was tuned to {\sc Hermes} data
to obtain the correct relative contribution from coherent and incoherent $\rho^0$ production \cite{mithesis}. 
This Monte Carlo generator also simulated the measured kinematic dependence of coherent $\rho^0$ production from 
heavier targets. The exclusive $\rho^0$ correction was 
found to be less than $1\%$ over the whole kinematic range, except at the highest $z$-value where it was about $10\%$. 
The contribution of vector-meson decay to the kaon sample was found to be negligible and no correction was included for this.
A systematic uncertainty due to {\sc Rich} hadron misidentification was estimated to be smaller than 1\%. 

The $p_t$-broadening for $\pi^+$, $\pi^-$, and $K^+$  was determined for four nuclei as a function of either $z$, $\nu$, $x$, or $Q^2$, while integrating over the other kinematic variables. 
The average kinematics of the $p_t$-broadening results are shown in Table~\ref{avtable}. The results reveal negligible correlations
among the variables, except for $x$ and $Q^2$ that are correlated due to the forward acceptance of the {\sc Hermes} spectrometer.

\begin{figure} 
  \begin{center}
    \includegraphics[width=0.85\linewidth]{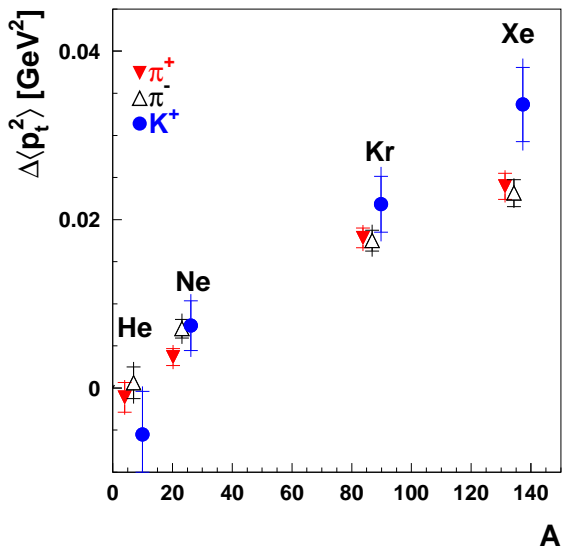}
    \caption{ The $p_t$-broadening for $\pi ^+$, $\pi ^-$, and $K^+$ mesons as a function  of atomic mass number A. The inner error bars 
represent the statistical uncertainties; the total bars represent the total uncertainty, obtained by adding
statistical and systematic uncertainties in quadrature. }
   \label{Adep}
  \end{center}
\end{figure}

The nuclear mass dependence of the $p_t$-broadening is presented in Fig.~\ref{Adep}. The broadening increases with mass number $A$. 
It is similar for the charged pions and seems to be systematically higher for positively charged kaons. 
The precision of the data does not allow a firm conclusion about the functional form of the increase of the data with A. There is no 
clear indication of a saturation of the $p_t$-broadening at large atomic mass numbers, 
supporting models which treat 
its origin in the partonic stage. Within such models, this behavior suggests that the color neutralization happens near the surface of the nucleus or 
outside for the average kinematics of this measurement \cite{accardi08}.

\begin{figure*} 
  \begin{center}
    \includegraphics[width=0.74\linewidth]{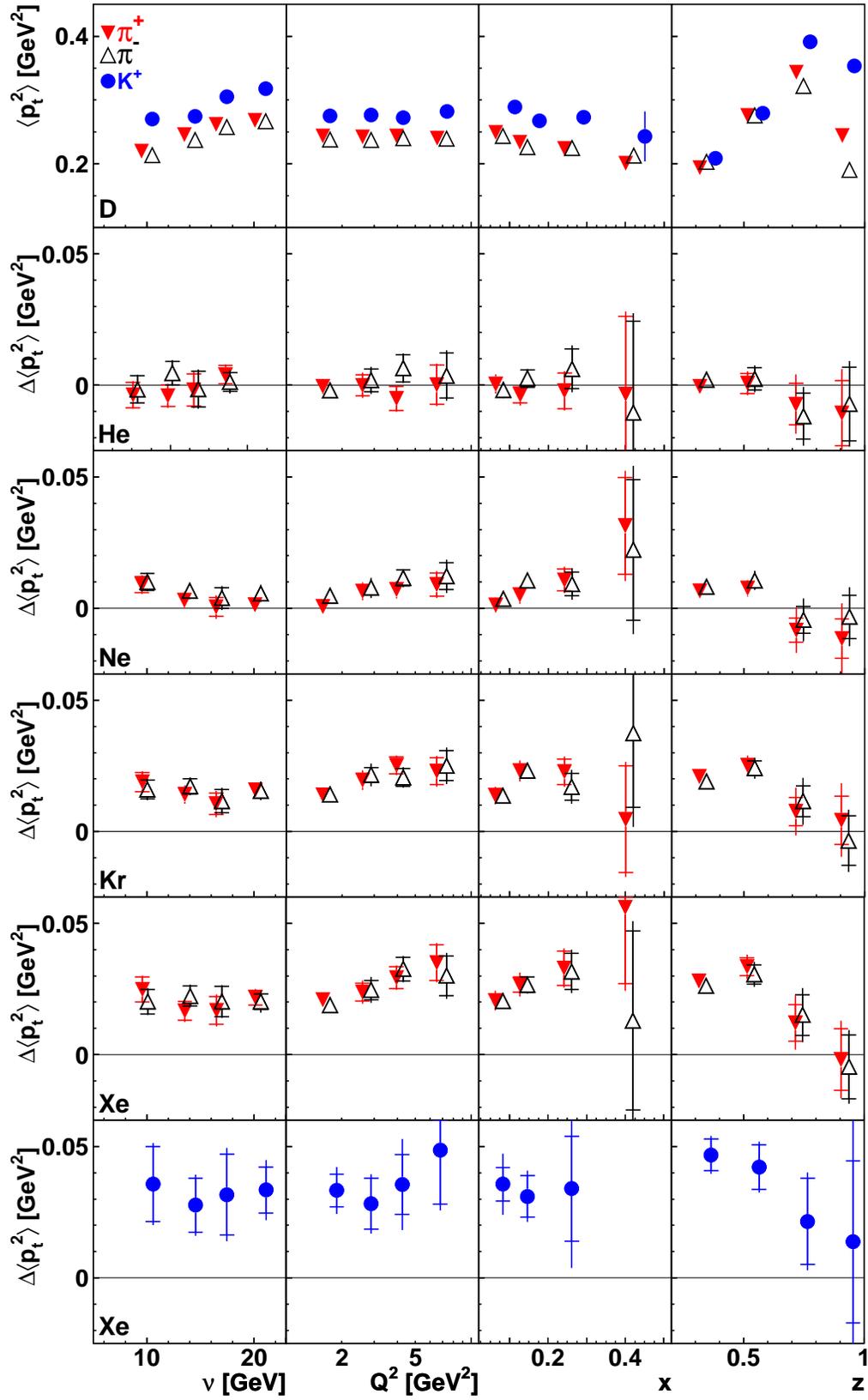}
    \caption{ From left to right, the $\nu$, $Q^2$, $x$, and $z$ dependence of $\langle p_t^2 \rangle$ for D (top row) and  $p_t$-broadening (remaining rows) 
for  $\pi^+$ and  $\pi^-$ produced on He, Ne, Kr, and Xe targets and for $K^+$ produced on a 
Xe target (bottom row). 
The inner error bars 
represent the statistical uncertainties; the total error bars represent the total uncertainty, evaluated as
the sum in quadrature of statistical and systematic uncertainties. 
}
   \label{ptbroad}
  \end{center}
\end{figure*}

The panels presented in Fig.~\ref{ptbroad} show $\langle p_t^2 \rangle$ for D (top row) and the $p_t$-broadening (remaining rows) as a function of either $\nu$, $Q^2$, $x$, and $z$ for $\pi^+$ or $\pi^-$ for the various nuclear targets. Since the uncertainties of the $K^+$ sample are rather large, only the results for  the Xe target are presented in the bottom row. 
The values of $\langle p_t^2 \rangle$ for D are between 0.2 and 0.4~GeV$^2$ while the $p_t$-broadening shows values from 0 up to 0.05~GeV$^2$. This means that $p_t$-broadening adds between 0 to 10\% to $\langle p_t^2 \rangle$. 
The data do not reveal a significant dependence on $\nu$ in the kinematic range covered. 

Since models that describe hadron formation in nuclei commonly connect formation length with $\nu$, the basically flat behavior 
in $\nu$  supports again the picture that color neutralization mainly happens at the surface (or outside) of the nucleus for the
{\sc Hermes} kinematics \cite{accardi08}. The effect slightly increases with $Q^2$
in contrast to the model calculation in Ref. \cite{kopel}, 
where a decrease of the broadening with $Q^2$ is predicted,  
and in agreement with the model calculation in Ref.~\cite{Domdey}.
The behavior as a function of $x$ is very similar to the $Q^2$ behavior, due to a strong correlation between $x$ and $Q^2$ in the {\sc Hermes} kinematics, hence it can not be excluded that the $Q^2$ dependence observed is actually an underlying x dependen
ce or both a $Q^2$ and $x$ dependence. The statistical precision of the data presented here do not
allow the study of the $Q^2$ and $x$ dependence separately, or any other two kinematic observables. 

The $p_t$-broadening is seen to vanish as $z$ approaches unity while the $\langle p_t^2 \rangle$ for 
D is 0.2 or higher in the highest $z$-bin.  Due to energy conservation 
the struck quark cannot have lost energy when $z = 1$, leaving no room for broadening apart from a possible modification of the primordial quark transverse momentum. The observed vanishing of the $\Delta \langle p_t^2\rangle _A^h$ at high values of $z$ indicates that there
 is no or little dependence of the primordial transverse momentum on the size of the nucleus. 
It also indicates that $p_t$-broadening is not due to elastic scattering of pre-hadrons or hadrons
already produced within the nuclear volume, as this would lead to substantial broadening even for values of $z$ very close to unity.

\begin{table}
\begin{center}
\begin{tabular}{|c||c|c|c|c|}
\hline
 & 
~$\langle \nu \rangle$[GeV]~  & ~$\langle Q^2 \rangle$[GeV$^2$]~ & 
~$\langle x \rangle$~  &~ $\langle z \rangle$~  \\
\hline
\hline
$\Delta \langle p_t^2\rangle$ vs. A &&&& \\
\hline
\hline
He & 13.7 & 2.4 & 0.101 & 0.42 \\
\hline
Ne & 13.8 & 2.4 & 0.101 & 0.42  \\
\hline
Kr & 14.0 & 2.4 & 0.100 & 0.41  \\
\hline
Xe & 14.0 & 2.4 & 0.099 & 0.41  \\
\hline
\hline
$\Delta \langle p_t^2\rangle$ vs. $\nu$ &&&&   \\
\hline
\hline
$\nu$-bin\# 1 & 8.0 & 2.1 & 0.141 & 0.49 \\
\hline
$\nu$-bin\# 2 & 11.9 & 2.5 & 0.111 & 0.43 \\
\hline
$\nu$-bin\# 3 & 14.7 & 2.6 & 0.096 & 0.40 \\
\hline
$\nu$-bin\# 4 & 18.5 & 2.4 & 0.073 & 0.37 \\
\hline
\hline
$\Delta \langle p_t^2\rangle$ vs. Q$^2$ &&&&   \\
\hline
\hline
$Q^2$-bin\# 1 & 13.7 & 1.4 & 0.063 & 0.42\\
\hline
$Q^2$-bin\# 2 & 14.0 & 2.5 & 0.105 & 0.41\\
\hline
$Q^2$-bin\# 3 & 14.4 & 3.9 & 0.153 & 0.40\\
\hline
$Q^2$-bin\# 4 & 14.6 & 6.5 & 0.248 & 0.39\\
\hline
\hline
$\Delta \langle p_t^2\rangle$ vs. $x$ &&&&  \\
\hline
\hline
$x$-bin\# 1 & 15.2 & 1.6 & 0.059 & 0.40\\
\hline
$x$-bin\# 2 & 12.3 & 3.0 & 0.131 & 0.42\\
\hline 
$x$-bin\# 3 & 11.5 & 5.5 & 0.254 & 0.42\\
\hline
$x$-bin\# 4 & 10.1 & 8.1 & 0.422 & 0.41\\
\hline
\hline
$\Delta \langle p_t^2\rangle$ vs. $z$ &&&&  \\
\hline
\hline
$z$-bin\# 1 & 14.5 & 2.4 & 0.097 & 0.32\\
\hline
$z$-bin\# 2 & 13.1 & 2.4 & 0.106 & 0.53\\
\hline 
$z$-bin\# 3 & 12.4 & 2.4 & 0.107 & 0.75\\
\hline
$z$-bin\# 4 & 10.8 & 2.3 & 0.115 & 0.94\\
\hline
\end{tabular}
\caption{Average kinematics for the ($\pi^+$) $p_t$-broadening results. The $\nu,~Q^2$, and $z$ kinematics are for the Xe target.}  
\label{avtable}
\end{center}
\end{table}

In summary, the first direct 
determination of $p_t$-broadening in semi-inclusive deep-inelastic scattering for charged pions 
and positively-charged kaons 
was performed on He, Ne, Kr, and Xe targets. 
The broadening was measured as a function of the atomic number $A$ and the
kinematic variables $\nu$, $Q^2$, $x$ or $z$. The broadening increases with $A$ and remains constant with $\nu$, suggesting that the effect is due to the ``partonic'' stage and that color neutralization happens near the surface or outside the nucleus.
 The effect increases with increasing $Q^2$ and $x$, and vanishes as $z$ approaches unity. 
 \begin{acknowledgments} 
We thank A. Accardi, B.Z. Kopeliovich, H.J. Pirner, and X.N. Wang for interesting and useful discussions. We gratefully acknowledge the {\sc Desy} management for its support, the staff at {\sc Desy} and the collaborating institutions for their significant
 effort, and our national funding agencies for financial support.

\end{acknowledgments}

%\special{
% ! userdict begin
% /bop-hook{
%   gsave 80 40 translate 58 rotate
%   /Times-Roman findfont
%   62 scalefont setfont
%   0 0 moveto
%   0.8 setgray
%   (Confidential, NOT for circulation!) show
%   grestore
%   } def
% end
% }

\end{document}